\documentclass[]{aa}
\usepackage{graphicx}

\begin{document}

\title{Electromagnetic radiation initiated by hadronic jets from microquasars 
in the ISM}

\author{V. Bosch-Ramon\inst{1} \and F. A. Aharonian\inst{2} 
\and J.~M. Paredes\inst{1}}

\institute{Departament d'Astronomia i Meteorologia, Universitat de Barcelona, Av. 
Diagonal 647, 08028 Barcelona, Spain
\and Max-Planck-Institut fur Kernphysik, Saupfercheckweg 1, Heidelberg, 69117, Germany}

\offprints{V. Bosch-Ramon \\ \email{vbosch@am.ub.es}}

\abstract{Microquasars are potential candidates to produce a non-negligible
fraction of the observed galactic cosmic rays. The protons accelerated at the jet 
termination shock interact with the interstellar medium and may produce detectable 
fluxes of extended emission at
different energy bands: high-energy and very high-energy $\gamma$-rays produced by neutral
pion-decay, synchrotron and
bremsstrahlung emission in a wide energy range generated by the secondary
electrons produced by charged pion-decay. We discuss the
association between this scenario and some of the unidentified
EGRET sources in the galactic plane.
\keywords{X-rays: binaries -- ISM: general -- gamma-rays: observations -- gamma-rays: theory}}

\maketitle

\section{Introduction} \label{Intro}

The high-energy $\gamma$-ray instrument EGRET detected about 170 sources that still remain 
unidentified. Among them, an important fraction are non-variable galactic sources. Several
astrophysical objects could explain the nature of this set of steady galactic $\gamma$-ray
sources: pulsars, supernova remnants, molecular clouds, etc. (see Torres et~al.
\cite{Torres03}, and also Nolan et~al. \cite{Nolan03}). However, new types of galactic sources
could be also the counterparts of this $\gamma$-ray steady emission detected by EGRET. In
particular, microquasars (MQs) have turned out to be likely sources of $\gamma$-rays (Paredes et~al.
\cite{Paredes00}), and they could be indirect sources of non-variable (at EGRET
timescales) $\gamma$-ray emission too. Indeed, these objects present powerful jets that show
persistent or transient activity and can carry huge amounts of kinetic energy during the life
of the MQ (millions of years for high-mass MQs and longer lifetimes for 
low-mass ones, Tauris \& Van den Heuvel \cite{Tau03}). Following the standard model for jet formation in MQs (i.e. Meier
\cite{Meier03}), the jets of microquasars transport a hadronic component extracted from the
accretion disk, which is formed by the matter accreted from the companion star. Once the jet
terminates, two different scenarios are possible: a smooth transition jet-interstellar medium
(ISM), where jet
protons would be released in the ISM with velocities similar to the original jet velocity, or
a shock-like interaction with the ISM, where jet protons would be accelerated by the Fermi
mechanism. If the first possibility occurs, these particles could be a minor but 
non-negligible part of the cosmic rays produced in our Galaxy at low energies (about GeV
energies). Moreover, in the second case, protons could reach higher energies, depending on the
conditions within the jet-ISM shock region (for an extended discussion, see Heinz \& Sunyaev
\cite{Heinz&sunyaev02}). From electrodynamical arguments (see, i.e, Aharonian
et~al. \cite{Aharonian02}), if the magnetic field and the size of the final part of the jet is
similar to that, for instance, considered in Paredes et~al. (\cite{Paredes02}), proton energies
of about 10$^5$ GeV may be achieved. Assuming that a non-thermal population of high energy
protons is released from the terminal part of a jet, these particles will diffuse through the
interstellar medium. At certain distances from the accelerator, the protons can interact with
high-density regions (i.e. molecular clouds), producing extended and steady emission at
different energies. Due to propagation effects, the outcomes from such interactions can differ
strongly depending on the age, the nature (impulsive or continuous) of the accelerator and the
distance between the accelerator and the cloud (see Aharonian \& Atoyan
\cite{Aharonian&atoyan96}). It should be noted that hadronic $\gamma$-rays could be produced 
at much smaller scales as well, e.g. at interaction of the relativistic jet of the compact
object  with the dense wind of the stellar companion, as suggested by Romero et al. 
(\cite{Romero03}).

In this work, we explore the possibility that MQs could initiate indirect and steady sources
of $\gamma$-rays and lower frequency radiation throughout interactions between high energy
protons released from their jets and nearby molecular clouds. We have developed a 
time-dependent model that calculates the broadband spectrum of the emission coming out from the pp
primary interactions as well as the emission produced by the secondary particles (electrons
and positrons) created during the first process. This model takes into  account the
propagation effects due to both energy-dependent diffusion and the pp interaction energy
losses of protons during their travel from the accelerator to the cloud. From this point, when
referring to an impulsive MQ, we mean a transient jet of MQ interacting with a cloud, and
by a continuous MQ, we mean the same but for a persistent jet of a MQ.

\section{MQ-cloud interactions and production of $\gamma$-rays} \label{MQ-cloud}

Jets of MQs must end somewhere, although the details of this are still unclear (Heinz
\cite{Heinz02}). Some MQs present persistent jets that emit at different energy bands up to
hundreds of AU with typical jet kinetic energy luminosities from 10$^{37}$ to
10$^{39}$~erg~s$^{-1}$  (LS~5039,  Rib{\'o} \cite{Ribo02}; SS~433, Marshall et~al.
\cite{Marshall02}, Spencer \cite{Spencer84}). Other members belonging to this type of galactic
jet sources present very powerful transient ejections that can extend 
larger distances and contain
several orders of magnitude more jet kinetic energy than the persistent cases (i.e. GRS~1915+105,
Mirabel \& Rodriguez \cite{Mirabel&rodriguez94}; Cygnus~X-3, Mart{\'{\i}} et~al.
\cite{Marti00}).  Finally, some MQs present extended structures at radio frequencies
(1E~1740.7$-$2942, Mirabel \& Rodr{\'{\i}}guez \cite{Mirabel&rodriguez99}) or at X-rays
(XTE~J1550$-$564, Corbel et~al. \cite{Corbel02},  Kaaret et~al. \cite{Kaaret03}) at scales of
0.1--1~pc, which could be related to acceleration of particles of a jet interacting with the
environment (i.e. Wang et~al. \cite{Wang03}).

Below, we initially assume that the jet has a significant population of protons. These protons
can be cold, or cooled during their propagation through the jet due to, i.e., adiabatic
losses.  They can be cold in the reference frame of the jet, but in the frame of the ISM their
energy  could be as large as $\Gamma m_{\rm p}c^2$, where $\Gamma$ is the Lorentz factor of
the jet and  $m_{\rm p}$ is the proton mass. However, they can also be accelerated in the
terminal part of the jet owing to shocks between the matter of the jet and the ISM. Therefore,
a population of non-thermal protons can extend up to very high energies, and these particles
are released since the low magnetic field strength is not enough to keep them within the
accelerator. Once the particles break free, they diffuse in the ISM with a diffusion speed 
depending on their energy and, close to or further from the releasing point, the protons interact
with higher density regions. The interaction between the high energy protons and the
interstellar hydrogen nuclei will produce charged and neutral pions ($\pi^-$, $\pi^+$ and
$\pi^0$); the first will decay to electrons and positrons and the second 
to photons. The primary radiation, $\pi^0$-decay photons, is in the $\gamma$-ray band,
but the secondary particles can produce significant fluxes of synchrotron (from radio
frequencies to X-rays) and bremsstrahlung emission  (from soft $\gamma$-rays to TeV range),
and generally with less efficiency, inverse Compton (IC) emission through interaction with the
ambient infrared (IR) photons.

\subsection{Evolution in time and distance of the proton energy distribution} \label{evolprot}

To calculate the emissivity of a source due to different mechanisms, we need to
know first what is the energy distribution of the high energy protons at 
different times and distances
from the acceleration site. Therefore, we have assumed that the (initial) energy distribution 
of the
accelerated protons follows a power-law plus a high-energy cutoff: 
\begin{equation}
f_{\rm p 0}(E_{\rm p})=K E_{\rm p}^{-p} \exp(-E_{\rm p}/E_{\rm p max}),
\label{eq:initfuncprot}
\end{equation}
where $K$ is the normalization
constant of the function (to compare with the total number of particles or total energy
contained by them), $E_{\rm p}$ is the proton energy and $E_{\rm p max}$ is the energy where 
the exponentical cut-off starts to dominate over the power-law. $E_{\rm p max}$ is 
determined by the efficiency of the accelerator and
will be treated here as a parameter, avoiding detailed treatments of the acceleration region.

To determine the energy distribution of protons in distance and time ($f_{\rm p}(E_{\rm p},R,t)$), 
we have used the solution found by Atoyan
et~al. (\cite{Atoyan95}) to the diffusion equation (Ginzburg \& Syrovatskii
\cite{Ginzburg&syrovatskii64}) in the spherically symmetric case:
\begin{equation}
\frac{\partial f_{\rm p}}{\partial t}=
\frac{D}{R^2}\frac{\partial}{\partial R}\left[R^2\frac{\partial f_{\rm p}}{\partial R}\right]+
\frac{\partial }{\partial E_{\rm p}}(Pf)+Q.
\label{eq:difeq}
\end{equation}
Here, $t$ is the source age, $R$ the distance to the source of protons, $D$ is the diffusion coefficient,
which is assumed as $D(E_{\rm p})\propto E_{\rm p}^{\chi}$, $P$ is the
continuous energy loss rate ($P=-dE_{\rm p}/dt$) and $Q$ is the source function, such that
$Q(E_{\rm p},R,t)=f_{\rm p 0}(E_{\rm p})\delta(\mathbf{R})\delta(t)$.  
$\delta(\mathbf{R})$ and $\delta(t)$ are delta functions in space and time, respectively.
To account for the energy losses
during the diffusive transport of protons, since the dominant cause of loss is nuclear
interactions, we have 
$P(E_{\rm p})=E_{\rm p}/\tau_{\rm pp}$, 
where $\tau_{\rm pp}\approx
6\times10^7(n/1~cm^{-3})^{-1}$~yr, and $n$ is the hydrogen density of the medium.
We assume an acceleration spectrum constant in time given by Eq.~(\ref{eq:initfuncprot}).

In our particular case of a power-law plus a high-energy cutoff injection spectrum, 
and a diffusion coefficient that depends on 
energy following a power-law, the general solution is reduced to:
\begin{eqnarray}
f_{\rm p}(E_{\rm p},R,t)\approx
\frac{K E_{\rm p}^{-p} \exp(-E_{\rm p}/E_{\rm p max})}{\pi^{3/2}R_{\rm dif}^3} \nonumber \\ 
\times \exp\left(-\frac{(p-1)t}{\tau_{\rm pp}}-\frac{R^2}{R_{\rm dif}^2}\right),
\label{eq:funcprot}
\end{eqnarray}
where $R_{\rm dif}$ is the diffusion radius and corresponds to the radius of the sphere
up to which the particles of energy $E_{\rm p}$ effectively propagate during the time $t$ after
they left the acceleration site,
\begin{equation}
R_{\rm dif}(E_{\rm p},t)=2\left(D(E_{\rm p})t\frac{\exp(t\chi/\tau_{\rm pp})-1}{t\chi/\tau_{\rm pp}} \right)^{1/2}.
\label{diff}
\end{equation}
$D(E_{\rm p})=D_{10}(E_{\rm p}/10$~GeV$)^{\chi}$, and $D_{10}$ is the diffusion coefficient
normalization constant in the ISM. For further details regarding diffusion of
particles in the ISM, see Aharonian \& Atoyan
(\cite{Aharonian&atoyan96}).

From Eq.~(\ref{eq:funcprot}) we can determine the proton energy distribution at any $t$ and
$R$, starting with an impulsively injected population of particles (Eq.~(\ref{eq:initfuncprot})).
To model the energy distribution of protons for a continuous MQ, we will need to integrate in
time from the initial to the present time, taking into account that the source function
represents the number of released particles  per energy and time units, and not the number of
released particles per energy unit, as in the impulsive case.

\subsection{Emission from interacting protons and hydrogen nuclei} \label{emissinter}

We calculate the radiative effects of diffusing high energy protons that 
penetrate a high density region of the ISM. We will assume that the channels of pp
interaction that yield either $\pi^0$, $\pi^-$ or $\pi^+$ are roughly equiprobable, thus the
functions that represent the photon, the electron and the positron energy distributions resulting
from the pion decay will be taken to be similar (in the relativistic regime of
secondary particles). 

The emissivity of the photons produced by $\pi^0$-decay ($q_{\gamma}$) at given $R$ and 
$t$ can be calculated through Eq.~(\ref{eq:phemis}):
\begin{eqnarray}
q_{\gamma}(E_{\gamma},R,t)=2 \int_{E_{\pi min}}^{E_{\pi max}} 
\frac{q_{\pi}(E_{\pi},R,t)}{\sqrt{ E_{\pi}^2-m_{\pi}^2c^4}}
dE_{\pi},
\label{eq:phemis}
\end{eqnarray}
where $E_{\gamma}$ and $E_{\pi}$ are the energy of the emitted photon and the decaying pion 
respectively, $E_{\pi min}=E_{\gamma}+m_{\pi}^2c^4/4E_{\gamma}$, $m_{\pi}$ is the pion mass and
$c$ is the speed of light. The neutral pion  emissivity from pp interaction ($q_{\pi}$) can be
found through Eq.~(\ref{eq:piemis}):
\begin{eqnarray}
q_{\pi}(E_{\pi},R,t)=c n_{\rm H}\int_{E_{\rm p min}}^{E_{\rm p max}} 
\delta(E_{\pi}-k_{\pi}E_{\rm kin}) \nonumber \\ \times
\sigma_{\rm pp}(E_{\rm p})n_{\rm p}f_{\rm p}(E_{\rm p},R,t)
dE_{\rm p} \nonumber \\ =
\frac{c n_{\rm H}}{k_{\pi}}\sigma_{\rm pp}(m_{\rm p}c^2+E_{\pi}/k_{\pi})
f_{\rm p}(m_{\rm p}c^2+E_{\pi}/k_{\pi},R,t),
\label{eq:piemis}
\end{eqnarray}
where $k_{\pi}$ is the mean fraction of the kinetic energy ($E_{\rm kin}=E_{\rm p}-m_{\rm
p}c^2$) of the proton transferred to the secondary pion 
per collision, $n_{\rm H}$ is the density of particles of the high density 
ISM region, and $\sigma_{\rm pp}(E_{\rm p})$ is the cross section of the pp interaction. 
This approach of emissivity calculation of $\pi^0$-decay $\gamma$-rays gives quite good accuracy
down to energies $\sim$1~GeV (see Aharonian \& Atoyan \cite{Aharonian00}).

Once we have obtained the emissivity, we need only to multiply by the total volume of
the interaction region ($V$) to find the total number of photons generated 
per energy unit and second. We suppose that the cloud is homogeneous. Also, its size
must be small enough compared to $R$ to assume the same energy distribution 
for the high energy
protons everywhere in the interaction region. 
Actually, one can assume that for a high density region with size smaller
than (but similar to) $R$, the changes in the proton spectrum 
are not significant enough to be taken into account at our level 
of approximation. Moreover, for the involved timescales, protons do not interact several times 
within the cloud, because the pp interaction cooling time is longer than the escape time 
in the proton energy range relevant here.
Then, the Spectral Energy Distribution (SED) of
the  emitted $\gamma$-rays due to the pp interaction within a cloud at a given $t$ (age of the
proton  source) and $R$ (distance from the proton source to the cloud) is:
\begin{eqnarray}
E_{\gamma}L(E_{\gamma})=E_{\gamma}^2 V q_{\gamma}(E_{\gamma},R,t).
\label{eq:piondec}
\end{eqnarray}
The number of electrons per time and energy units produced within a cloud at a certain $R$ 
and $t$ ($Q_{\rm e}(E_{\rm e},R,t)$) presents a similar distribution, where $E_{\rm e}$ 
is the electron energy.

\subsection{Secondary particles and their emission} \label{secpart}

We have introduced bremsstrahlung, synchrotron and IC energy losses for the secondary leptons
in order to get the proper electron energy distribution. Also, we have assumed that leptons
remain within the cloud after their creation. This is a reasonable assumption as a first 
order approach for electrons 
emitting via synchrotron process at frequencies above 10~GHz, taking into
account that the cooling time of those electrons within the cloud is shorter than their escape
time, assuming a diffusion for electrons inside similar to the one for protons outside. 
We note also that for sources younger than $\sim10^4$~yr, secondaries emitting roughly below 
10~GHz have not had enough time to escape yet.  
We will not study emission at lower energies, since this requires 
a more complex approximation,
which will be performed in future work.
We have determined the dependence on time of the electron
energy  adopting quite reasonable assumptions: homogeneous and isotropic velocity
distribution of electrons, a homogeneous distribution of hydrogen nuclei, and constant magnetic
and radiation fields ($B$ and $U$ respectively). Regarding radiation fields, we have taken into
account the 2.7~K photon galactic  background plus an additional 50~K photon field considered
to be produced inside the cloud (see Table~\ref{constants}). These IC losses, although
unavoidable,  are not significant unless very high density IR fields are present, 
which has not
been excluded (see Sect.~\ref{IC}). The electron energy is given by:
\begin{eqnarray}
E_{\rm e}(t)=\frac{a_0 n_{\rm H} E_{\rm e 0}}
{(a_0 n_{\rm H}+E_{\rm e 0}C)\exp[a_0n_{\rm H}(t-t_0)]
-E_{\rm e 0}C},
\label{eq:evol} 
\end{eqnarray}
where $C=(a_{\rm s}B^2+a_{\rm c}U)$, $a_0=8.1\times10^{-16}$, $a_{\rm s}=2.37\times10^{-3}$
and  $a_{\rm c}=3.97\times10^{-2}$ in cgs units. Here, $a_0$ is related to the bremsstrahlung 
losses, and $a_{\rm s}$ and $a_{\rm c}$ are associated with the synchrotron and the IC losses
(Thomson regime) respectively. $t_0$ is the initial time, corresponding to an electron energy
$E_{\rm e 0}$. 

From $E_{\rm e}(t)$ and the continuity equation we can predict how the 
injected secondary particles, following an initial $Q_{\rm e}$, will change their energy 
distribution in time. 
Knowing how an injected $Q_{\rm e}$ evolves in time, we integrate the different evolved
values of $Q_{\rm e}$  from the initial time to the present moment. Doing so, we get the total
number of electrons per energy unit at any $t$ and $R$: $N_{\rm e}(E_{\rm e},R,t)$.  

We can compute $N_{\rm e}(E_{\rm e},R,t)$ using numerical calculations for any value of
$E_{\rm e}$, $R$ and $t$. Knowing the previous function, the magnetic and radiation fields,
and the density of the cloud, one can easily calculate the emission from these secondary
particles produced via synchrotron radiation, bremsstrahlung and IC scattering (i.e.
Blumenthal \& Gould \cite{Bg}). 

\section{Performance of the model for different cases} \label{Applying}

\begin{table*}
\begin{center}
\caption[]{Adopted parameter values.}
\begin{tabular}{l c c c c c}
\noalign{\smallskip} 
\hline 
\hline 
\noalign{\smallskip} Parameter & Symbol & Value 
\cr \noalign{\smallskip} \hline \noalign{\smallskip} 
Diffusion coefficient normalization constant & $D_{10}$ & $10^{27}$~cm$^2$~s$^{-1}$
\cr Diffusion power-law index & $\chi$ & 0.5
\cr ISM medium density & $n$ & 0.1~cm$^{-3}$
\cr High density ISM region/cloud density & $n_{\rm H}$ & 10$^4$~cm$^{-3}$
\cr Mass of the high density ISM region/cloud & $M$ & $5\times10^3~M_{\odot}$
\cr Magnetic field within the cloud & $B$ & $5\times10^{-4}$~G
\cr IR radiation energy density within the cloud & $U$ & 10~eV~cm$^{-3}$
\cr Planckian grey body temperature (IR) & $T$ & 50~K
\cr Power-law index of the high energy protons & $p$ & 2
\cr Cut-off energy of the high energy protons & $E_{\rm p max}$ & 10$^5$~GeV
\cr Kinetic energy luminosity for accelerated protons in the continuos MQ & $L_{\rm k}$ & 10$^{37}$~erg~s$^{-1}$
\cr Kinetic energy for accelerated protons in the impulsive MQ & $E_{\rm k}$ & 10$^{48}$~erg
\cr \noalign{\smallskip} \hline
\end{tabular}
\label{constants}
\end{center}
\end{table*}

We go through different cases that might be relevant. The adopted parameter values are presented
in Table~\ref{constants}, although some of them could change depending on the particular
scenario. We have computed the emission produced by four different radiative processes:
$\pi^0$-decay, synchrotron radiation, bremsstrahlung and IC scattering. For the impulsive case,
we have considered a very active source, $L_{\rm k}\sim $10$^{39}$~erg~s$^{-1}$, but operating a
short period of time, $\sim 30$~yr, which implies a $E_{\rm k}\sim$10$^{48}$~erg (transfered to
the accelerated protons). We note that this value for the kinetic energy luminosity of
protons is average during the impulsive phase, and it takes into account periods
of less activity and recurrent super-Eddington accretion bursts within these
$30$~yr. For a continuous MQ, we have adopted moderate kinetic energy luminosities for the jet
(transfered to the accelerated protons) of about 10$^{37}$~erg~s$^{-1}$ (see
Table~\ref{constants}). 

In the case of impulsive MQ, as is mentioned in Sect.~\ref{MQ-cloud} for the case of SS~433, we have a
persistent jet presenting a  kinetic energy luminosity similar to or even more than $L_{\rm
k}\sim 10^{39}$~erg~s$^{-1}$  (Marshall et~al. \cite{Marshall02}, Spencer \cite{Spencer84}). It
is a super-Eddington accretion system (as can be seen from the huge amount of jet kinetic energy
), but the ejection appears to be persistent. It could also be the case that a microquasar
suffering recurrent or periodic outbursts delivered, on average, the amount of proton kinetic
energy assumed here. For instance, GRS~1915+105, which can easily reach super-Eddington accretion
rates, presents large outbursts like the one in 1994 (Mirabel \& Rodriguez \cite{Mirabel&rodriguez94}), with
kinetic energy luminosities close to 10$^{41}$~erg~s$^{-1}$ (Gliozzi  et~al. \cite{Gliozzi99}).
Otherwise, for a continuous MQ, assuming a proton acceleration efficiency of 10\% and a jet
carrying 20\% of the accretion power, an accretion rate of 10$^{-8}~M_{\odot}$~yr$^{-1}$ is 
required. Therefore, the energetic assumptions of this work are not rare for microquasars. 

Regarding $p$, we have adopted a value of 2. Cosmic ray spectrum
is assumed to be injected with power-law indices of around 2.3. However, from indirect spectral
considerations of radiation emitted from the jets of MQs (i.e. their electron spectrum inferred
from synchrotron emission), it seems more suitable in that case to take a slightly harder value.
The different explored situations are presented in the next subsections. For each case, 
the relevant paramater values are indicated in italic.

\subsection{Proton flux evolution} \label{Protspecevol}

{\it $R$=10~pc; $t$=100, 10000~yr.}

\begin{figure}
\resizebox{\hsize}{!}{\includegraphics{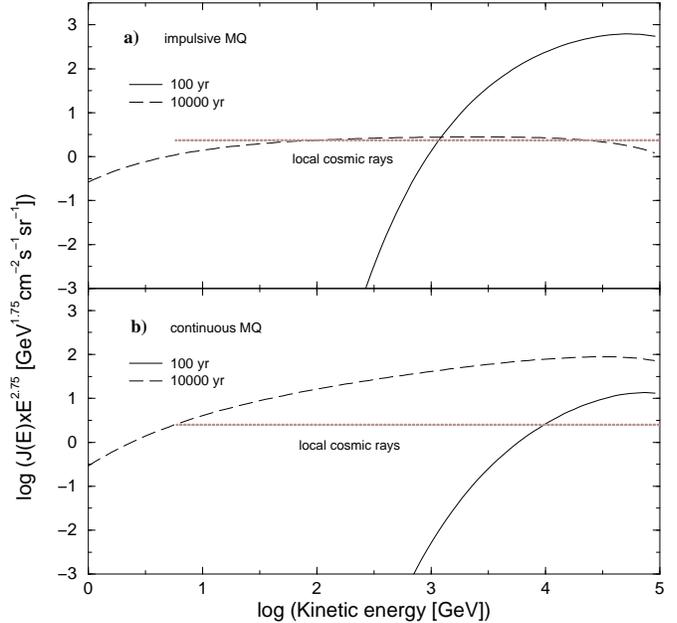}} \caption{Spectrum of protons released
from a source at $R$=10~pc after $t$=100~yr (solid line) and 10000~yr (long-dashed line). The
dotted horizontal line shows the flux of cosmic ray protons observed near the Earth.
\textbf{a)} For an impulsive MQ. \textbf{b)} For a continuous MQ.}
\label{protesp}
\end{figure}

We have computed the proton spectra ($J(E_{\rm p})\times E^{2.75}$) for an impulsive and a
continuous MQ together with the cosmic ray spectrum. The results are plotted in Fig.~\ref{protesp}.
We note a harder spectrum and more particles for a continuous MQ compared
to an impulsive MQ at older ages. This is due to the continuous particle injection from the jet
of the MQ in the first case.

\subsection{Proton and electron energy distributions} \label{Prontandelec}

{\it $R$=10~pc; $t$=100, 1000~yr.}

\begin{figure}
\resizebox{\hsize}{!}{\includegraphics{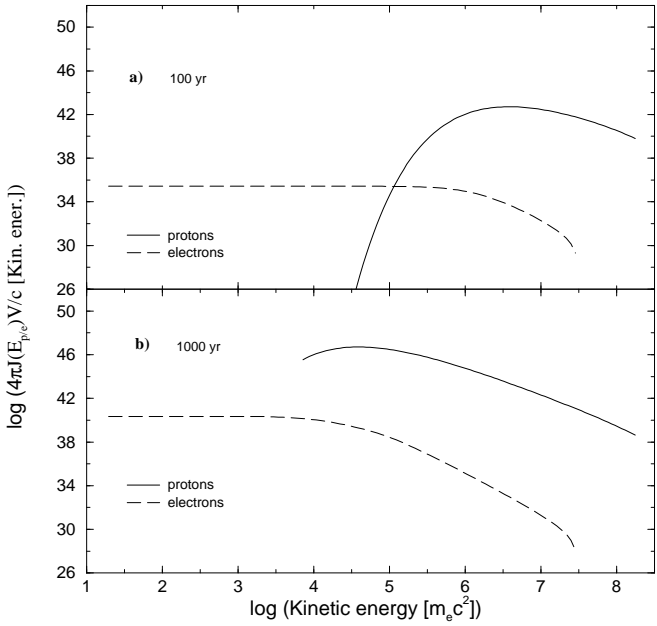}} \caption{Proton and electron energy 
distributions (solid and long-dashed lines respectively) for an impulsive MQ at 
$R$=10~pc. The kinetic energy axis units are m$_{\rm e}$c$^2$.
\textbf{a)} The source age is $t$=100~yr. \textbf{b)} The source age is 1000~yr.}
\label{prel1}
\end{figure}

\begin{figure}
\resizebox{\hsize}{!}{\includegraphics{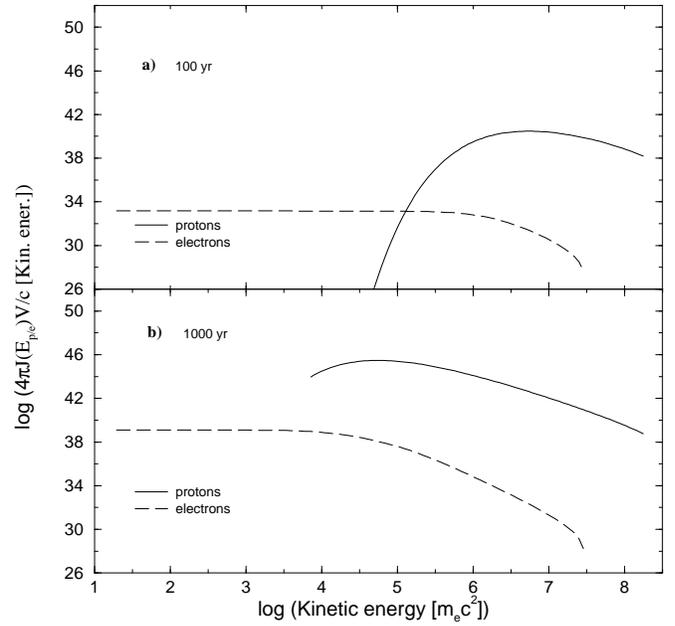}} \caption{
The same as in Fig.~\ref{prel1} but for a continuous MQ 
(recall: \textbf{a)} $t$=100~yr, \textbf{b)} $t$=1000~yr).}
\label{prel2}
\end{figure}

We have computed the proton and the electron energy distributions ($J(E_{\rm p/e})\times
\frac{4\pi}{c}V$) within a cloud, plotting them together in the same graph to demonstrate their
evolution with time. The impulsive and the continuous cases are shown in
Figs.~\ref{prel1}~and~\ref{prel2} respectively. We note that, from the diffusion radius
(Eq.~(\ref{diff})) and the time-dependent electron energy  (Eq.~(\ref{eq:evol})), we can
recognize the peaks and turning points of both distribution functions at those energies that
correspond roughly to the protons and electrons that dominate at that time. Another interesting
feature of Figs.~\ref{prel1}~and~\ref{prel2} is that the number of protons and electrons per
volume and energy units in the continuous case moves closer to the number in the impulsive case
as $t$ is increased, since for the continuous source the protons and electrons accumulate ({\it
accumulate} for protons) in the cloud with a continuous proton injection along the MQ lifetime,
whereas in the impulsive MQ protons do not {\it accumulate} and secondary particles are
injected  with a changing rate. For longer periods of time, the impulsive MQ emission will
eventually disappear, but the continuous MQ will be emitting at least during its whole life.
The secondary particles accumulate also in the cloud for the impulsive MQ, but there is no
injection of fresh particles at the highest energies, so the spectrum becomes steeper than in
the continuous MQ. Below a certain energy, the electron energy distribution is  completely
flat. This comes from both the lower energy cutoff of the proton energy distribution due to 
the dependence on energy of the diffusion in the ISM and the characteristics of the pp
interaction and production of secondary particles.

\subsection{Overall spectrum at different ages} \label{sourcage}

{\it $R$=10~pc; $t$=100, 1000, 10000~yr.}

\begin{figure}
\resizebox{\hsize}{!}{\includegraphics{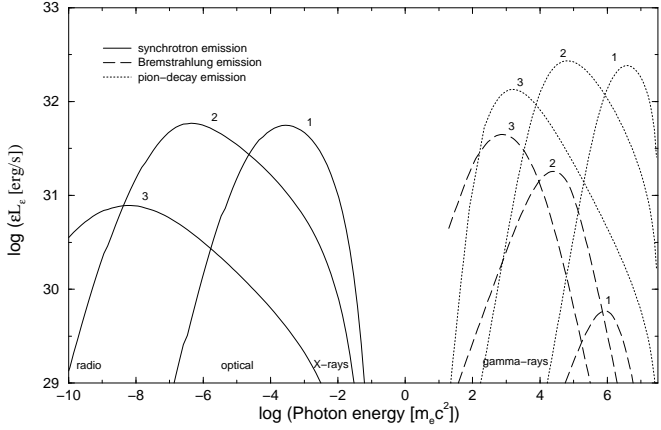}} \caption{The SED for an 
impulsive MQ for $R$=10~pc at three different epochs: for 100~yr (curve 1), for 1000~yr (2),  
and for 10000~yr (3). The photon energy axis units in 
this and following plots are m$_{\rm e}$c$^2$.} 
\label{specagei} 
\end{figure}

\begin{figure}
\resizebox{\hsize}{!}{\includegraphics{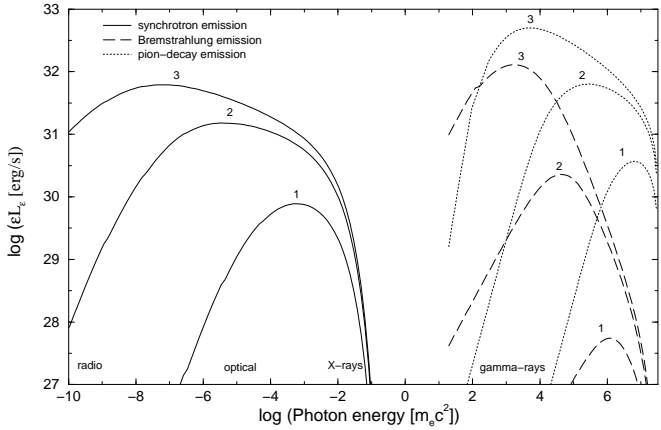}} \caption{The same as in Fig.~\ref{specagei}
but for a continuous MQ (recall: $t$=100~yr (1), 1000~yr (2), and 10000~yr (3)).}
\label{specagec}
\end{figure}

We have computed the SED from radio to $\gamma$-ray energies at
different $t$. Both the impulsive and the continuous MQ cases are presented in
Figs.~\ref{specagei}~and~\ref{specagec}, respectively. The luminosity peaks of primary and
secondary particle emission are shifted to lower energies as incoming primary particles
have less energy and secondary electrons lose their energy. Also, the secondaries accumulate
within the cloud (assuming they do not escape), decreasing the difference between the fluxes from 
the $\pi^0$-decay and bremsstrahlung channels.

The comparison of Figs.~\ref{specagei}~and~\ref{specagec} shows different spectral
features. The spectrum for an impulsive MQ at low and high energies is steeper and presents
lower energy luminosity peaks than the one for a continuous source. Also, the total emission
increases first and decreases later for the first case, increasing asymptotically
for the second case. This is due to the lack of continuous injection of new protons at high
energies and the reduction of the number of protons at different energies with time for an
impulsive MQ. In the continuous source, since protons are continuously arriving at the cloud
following the same energy distribution, there are cumulative effects on radiation. The previous
issues make the impulsive case a good candidate to be detected at {\it early} stages of the source
age (100--1000~yr) and at higher energies (i.e. detectable by the new generation of ground-based 
Cherenkov telescopes),
and a good case to determine the energy of the progenitor particles of the emission (due to the
weaker cumulative effects of the spectrum). Otherwise, the continuous case should be easier to
detect at later stages of the source age ($>10000$~yr) and at lower energies (i.e.
detectable by EGRET), but the cumulative spectral effects would give a bias in the extrapolation
of the typical particle energies from the peaks in the spectrum\footnote{Actually, for
secondaries, what should be studied  is the turning point in the spectrum due to energy losses.}.

\subsection{Spectra for different diffusive behaviors and proton origins} \label{difspec}

{\it $R$=10~pc (continuous MQ); steady regime; $E_{\rm pmax}\sim 10^6$~GeV; $p=2.75$ (cosmic rays)
and $=2$ (continuous MQ).}

\begin{figure}
\resizebox{\hsize}{!}{\includegraphics{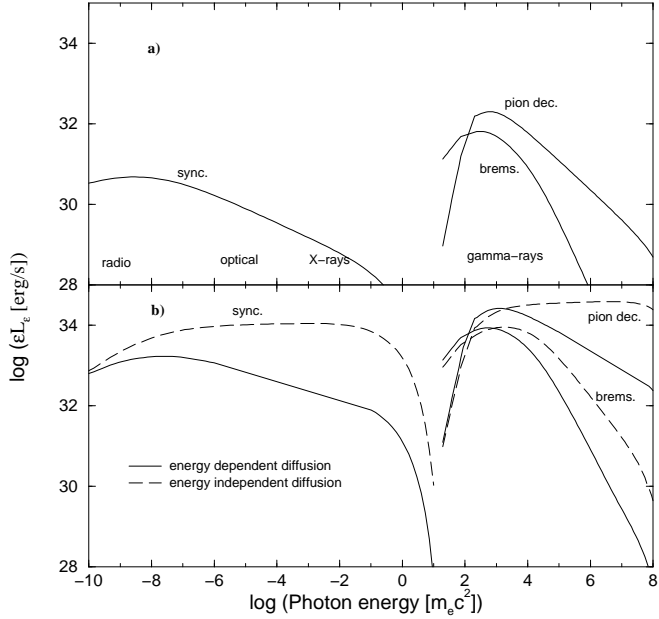}} \caption{ 
\textbf{a)} SED from the galactic cosmic rays interacting with a cloud. 
The power-law index for proton energy distribution has been taken to be 2.75, as is 
found near the Earth. Diffusion does not depend on energy. \textbf{b)} 
SED from a continuous MQ: in one case the diffusion is energy dependent (solid line), 
in the other case it is not (long-dashed line).}
\label{comsp}
\end{figure}

In this subsection, we have computed the overall spectrum assuming that the relativistic
protons belong to the sea of cosmic rays. We have computed also the spectra for two continuous
MQ cases: an energy-dependent diffusive case and a energy-independent diffusive case. Since
the cosmic ray sea can be considered to follow a homogeneous energy distribution in space and
time, for this case we have assumed that there is no dependence on energy in the diffusive
process, as well as an accelerator age long enough to reach the steady regime. In the second
situation, it is possible to compare the energy-dependent diffusion spectrum with the energy
independent diffusion spectrum, taking the initial energy distribution of protons to be
exactly the same (power-law index, kinetic energy luminosity of the jet, etc...). Again, the
age has been taken long enough to reach the steady regime for relativistic protons within the
cloud. In Fig.~\ref{comsp}, upper panel, the cosmic ray case is shown, the two continuous MQ
cases being presented in the lower panel. Since galactic cosmic rays can reach higher energies
than the adopted maximum energies for protons, we have now taken $E_{\rm p max}\sim
10^6$~GeV in the three SED to compare them properly.

Fig.~\ref{comsp} demonstrates the importance of energy-dependent diffusion in obtaining the
correct spectral shape. Without energy dependence in the diffusive process, the fluxes grow at
all the wavelengths in such a way that there is neither spectral steepening nor emission peak
shifting with time. Also, comparing the continuous MQ cases with the cosmic ray one, there is
an important difference in the spectral slope between the case with no energy dependence in
diffusion and the cosmic ray case, but not between the one with energy dependence in diffusion
and the cosmic ray case. In the context of our model, it means that the origin of the cosmic
ray sea is a group of sources which are far enough from their corresponding interacting clouds
to suffer steepening in the proton energy distribution, following these protons initially a
power-law of index $\sim$2 (like the typical value adopted in this work).

\subsection{Overall spectrum at different distances} \label{sourcdis}

{\it $R$=10, 30, 100~pc; $t=1000$~yr.}

\begin{figure}
\resizebox{\hsize}{!}{\includegraphics{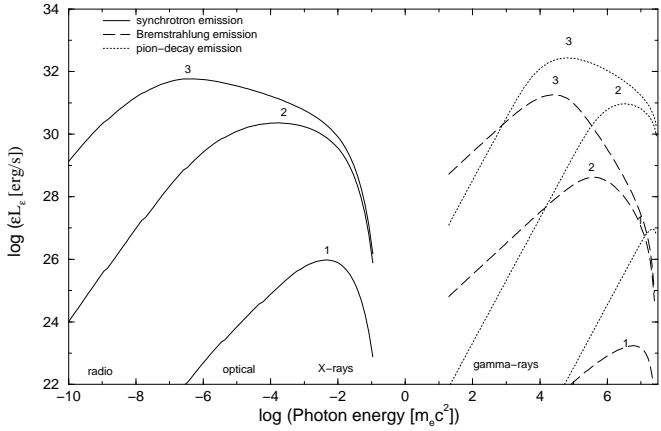}} \caption{
The SED for an
impulsive MQ for $t$=1000~yr at three different $R$: for 100~pc (curve 1), for 30~pc (2), and for 10~pc
(3).}
\label{specdisi}
\end{figure}

\begin{figure}
\resizebox{\hsize}{!}{\includegraphics{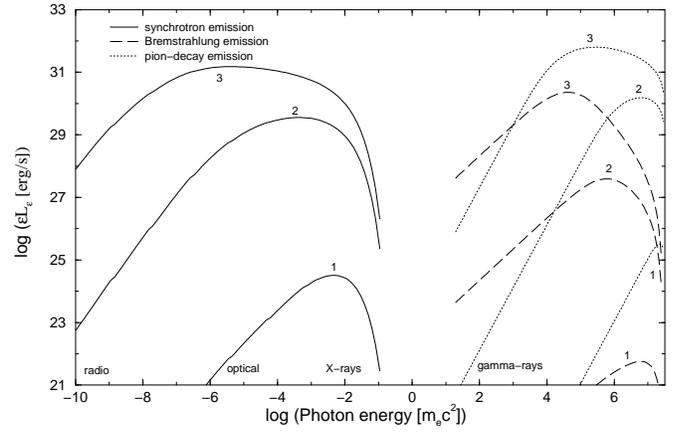}} \caption{The same as in Fig.~\ref{specdisi}
but for a continuous MQ (recall: $R$=100~pc (1), 30~pc (2), and 10~pc (3)).}
\label{specdisc}
\end{figure}

In Figs.~\ref{specdisi}~and~\ref{specdisc}, we present the computed SED from radio to
$\gamma$-ray energies of an impulsive and a continuous MQ, respectively. Due to the effects
imposed by the propagation of particles in the ISM, there is a relationship between time and
distance (see Eq.~(\ref{diff})). This fact implies that
Figs.~\ref{specdisi}~and~\ref{specdisc} will present similar features to 
Figs.~\ref{specagei}~and~\ref{specagec}, like the decreasing difference in luminosity between
$\pi^0$-decay and bremsstrahlung emission and the shift of the emission peaks to lower
energies. 

\subsection{Low and high magnetic fields} \label{magnetic}

{\it $R$=10; $t=100$, 10000~yr; $B$=10$^{-3}$, 10$^{-5}$~G.}

\begin{figure}
\resizebox{\hsize}{!}{\includegraphics{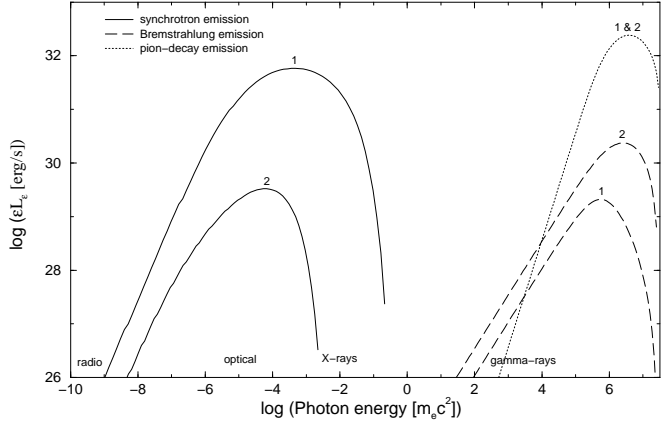}} \caption{
SED of an impulsive MQ taking $R$=10~pc, $t$=100~yr, and two magnetic field strength 
$B$=10$^{-3}~G$ (1) and 10$^{-5}~G$ (2).}
\label{magn1}
\end{figure}

\begin{figure}
\resizebox{\hsize}{!}{\includegraphics{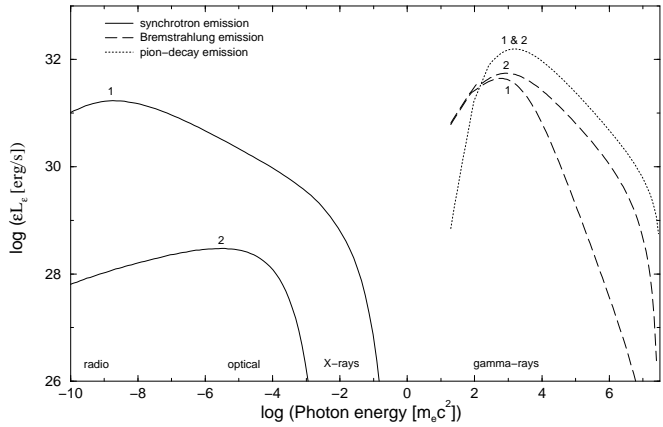}} \caption{
The same as in Fig.~\ref{magn1}, but for a $t$=10000~yr 
($B$=10$^{-3}~G$ (1) and 10$^{-5}~G$ (2)).}
\label{magn2}
\end{figure}

We have calculated the SED of an impulsive MQ taking a low and a high value for $B$. For a young
source case, plotted in Fig.~\ref{magn1}, and a low $B$, owing to the small loss rate of
secondary particles, the position in energy of the maximum of emission for the secondary
particles mainly depends on the typical energy of the arriving protons at that $R$ and
$t$. For a young source and a high $B$ (same Fig.~\ref{magn1}), we recover the strong dependence
of the secondary particles on time due to the losses. Otherwise, for an older source case,
plotted in Fig.~\ref{magn2}, and a low $B$, the peak for bremsstrahlung shows at what energy
the electrons are concentrated (roughly the same as protons) although the synchrotron emission
peaks at higher energies. For an older source and a high $B$ (same Fig.~\ref{magn2}), the
computed SED again shows strong dependence of the secondary particles on time due to the
losses, presenting a softer spectral slope than in the case of a small $B$ at the same age. In
all the previous situations it can be seen how bremsstrahlung and synchrotron are processes
more or less important to each other depending on the synchrotron losses: for
high $B$, synchrotron dominates, for low $B$, bremsstrahlung dominates. Of course, the magnetic
field does not affect $\pi^0$-decay emission.

In previous subsections the IC component has not been shown. This is due to the low
emission fluxes produced via IC scattering between secondaries and IR photons within the cloud
for the adopted density of such photons (see Table.~\ref{constants}), much lower than the ones
predicted for the other components (synchrotron, bremstrahlung and pion-decay). Nevertheless, in
the next subsection, a study about the importance of the IC radiation is presented.

\subsection{On the importance of the IC radiation} \label{IC}

{\it $R$=10; $t=1000$; $U$=1, 10$^4$~eV~cm$^{-3}$.}

\begin{figure}
\resizebox{\hsize}{!}{\includegraphics{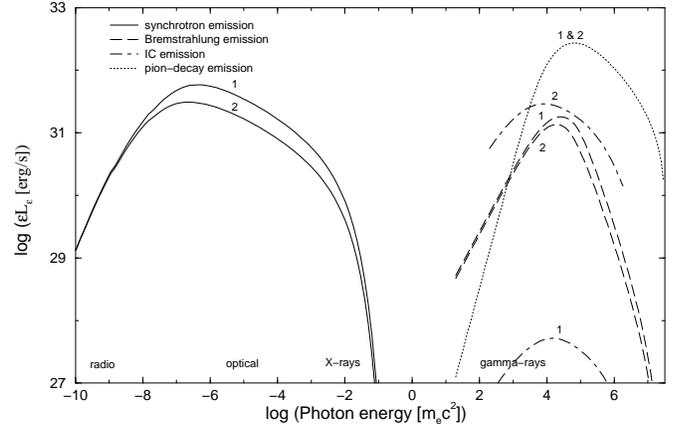}} \caption{SED of an impulsive source
at $R$=10~pc after $t$=1000~yr. We have plotted both the cases for $U=1$~eV~cm$^{-3}$ (1) 
and $U=10000$~eV~cm$^{-3}$ (2).}
\label{ICf}
\end{figure}

In Fig.~\ref{ICf} we show the importance of the IC emission comparing with the radiation from
other components (synchrotron, bremsstrahlung and $\pi^0$-decay processes) depending on the
energy density of the infrared seed photon field. We recall that the IR spectrum has been
taken to be a Planckian grey body, with a temperature of about 50~K.  From Fig.~\ref{ICf}, it
is seen that a high density IR photon field ($\geq 10^4$~eV~cm$^{-3}$) is necessary within the
cloud to have a relevant IC component in radiation. Thus, for clouds with standard IR photon
energy densities, the dominant source of electron energy losses would be the magnetic field,
and for clouds with IR photon energy densities about four orders of magnitude higher than the
standard value, the dominant source of electron energy losses would be IC scattering, with IC
emission levels similar to those generated by the bremsstrahlung process in the same energy
range. However, IR field densities of 10$^4$~eV~cm$^{-3}$ are too high, such a situation being
very unlikely. Of course, $\pi^0$-decay emission is not affected by the IR photon field
density. It is interesting to note the small decrease of synchrotron and bremsstrahlung
emission due to the moderate increase of losses because of the higher radiation field density,
compared to the strong increase of IC emission. It seems very unlikely that IC losses could
be important enough to control by themselves the radiative processes and the evolution of
secondary particles in a reasonable scenario.

\subsection{Continuous MQ with a strong outburst} \label{Contburst}

{\it $R$=10~pc; $t=10^5$~yr (continuous MQ) and $t=200$~yr (impulsive MQ).}

\begin{figure}
\resizebox{\hsize}{!}{\includegraphics{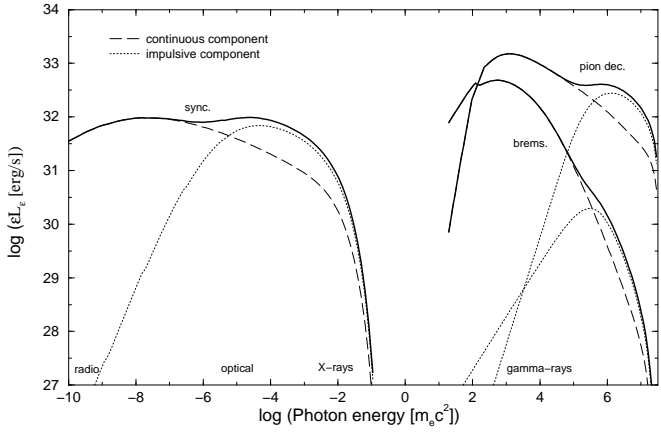}} \caption{
SED of a complex source: a continuous MQ (long-dashed line) 
plus an impulsive phase (dotted line). $t$=10$^5$~yr for the first one
and $t$=200~yr for the second one, with $R$=10~pc. The added spectrum is presented as a thick
solid line. The kinetic energy luminosity for the continuous MQ has been taken to be 
10$^{37}$~erg~s$^{-1}$, and the kinetic energy for the impulsive MQ has been taken to be 
10$^{48}$~erg.}
\label{contimp}
\end{figure}

Now we combine two kind of sources, an impulsive MQ and a continuous MQ, to obtain the complex
SED that is presented in Fig.~\ref{contimp}. To find clouds presenting emission with different
origins is quite probable. Taking into account the long-term variable nature of microquasars
(the short-term is not taken into account here because it does not affect large-scale objects
like clouds), we can find different periods of activity such as persistent and impulsive phases.
Therefore, a spectrum like the one shown in Fig.~\ref{contimp} might be quite typical. Due to
obvious differences in evolution for both types of emission, that originating in the
continuous phase and that originating in the impulsive phase, we find that the spectral
shape presents two different peaks and complicated slopes, being detectable 
at radio, X-ray, EGRET and TeV enegies.

\subsection{The case of two clouds} \label{physicalinf}

{\it $R$=5~pc and $R$=50~pc, with $V\sim 3.4$ and 3400~pc$^3$,
respectively.}

\begin{figure}[T]
\resizebox{\hsize}{!}{\includegraphics{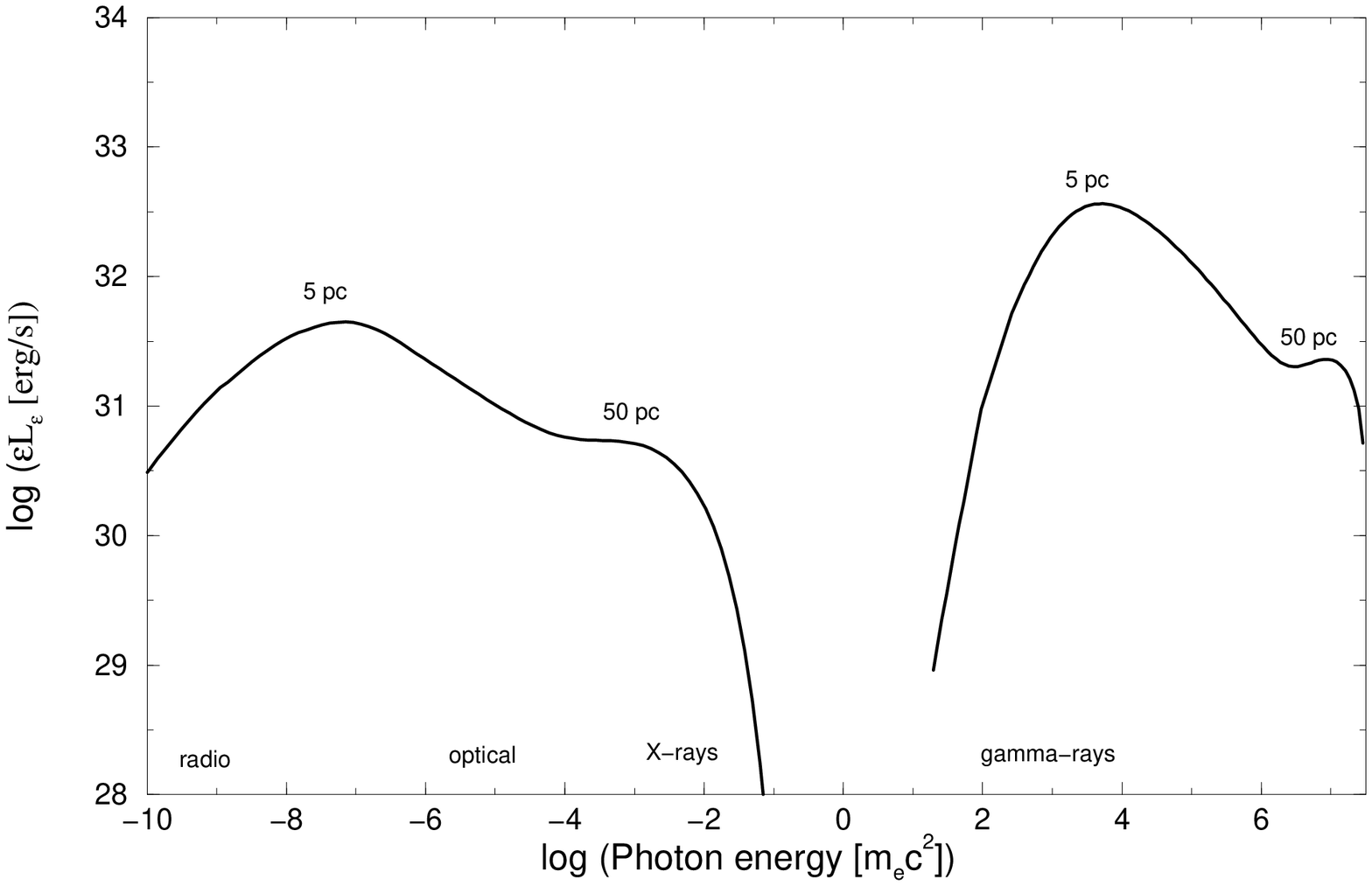}} \caption{SED of the observed spectrum from two
clouds interacting with an impulsive MQ at distances 5 and 50~pc .}
\label{twoclou}
\end{figure}

In Fig.~\ref{twoclou} we have computed the spectrum that would be observed if there were two
clouds at different known distances from the origin of the particles. We assume an impulsive
MQ as being more clear than the continuous one to determine its physical parameter values due
to the reduced particle accumulation effects. All the other parameters except the distance
from the clouds to the MQ would be unknown a priori, and our task is to estimate them
from the observed spectrum using our model. 

From the characteristic frequency of the synchrotron emission ($\nu_{\rm c}=6.3\times10^{18} B
E_{\rm e}^2$) and the maximum energy reached by both the synchrotron and the $\pi^0$-decay
emissions, we can roughly determine the magnetic field within the clouds ($5\times 10^{-4}$~G)
and the electron energy at the peaks and turning points. In addition,
Eqs.~\ref{diff}~and~\ref{eq:evol} can help to estimate the time of travel for the first particles
from the MQ to the cloud and the time during which the secondary particles have evolved
(the total age, $t$=1000~yr). From previous parameters, we estimate the diffusion
coefficient constant ($D_{10}=10^{27}$~cm$^2$~s$^{-1}$). The spectral shape can also give
information about the proton energy distribution. Knowing the previous set of parameter values
and the volume of the cloud, it could be possible to find the remaining parameters: $n_{\rm H}$
and the total kinetic energy of the released protons.  

\section{MQ, clouds and non variable $\gamma$-ray sources} \label{nonvar}

\begin{figure}
\resizebox{\hsize}{!}{\includegraphics{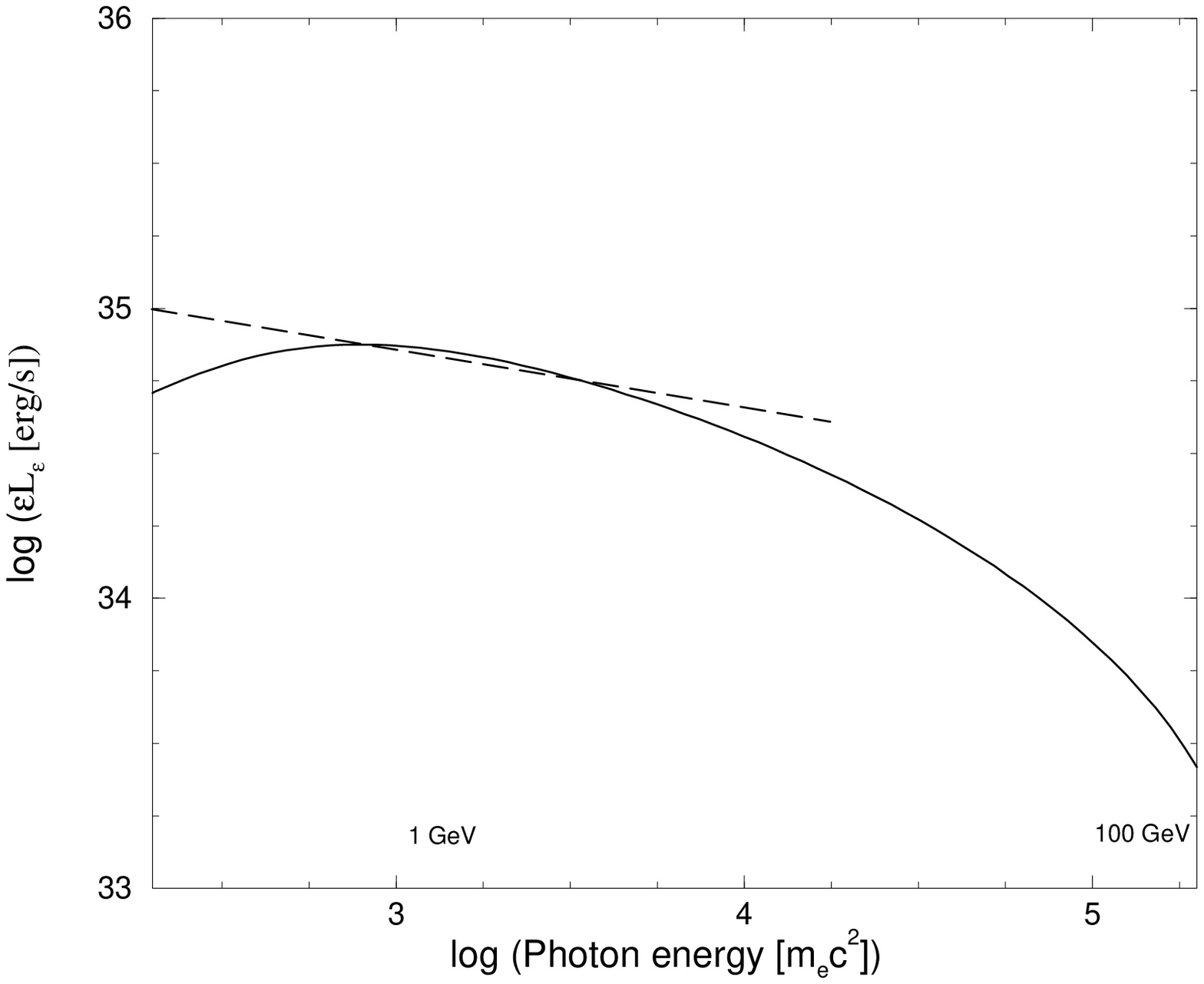}} \caption{
SED of a continuous MQ at $t$=10$^6$~yr and $R$=10~pc reproducing a typical 
GeV source (solid line). The value
of the cloud mass is 10$^5$~M$_{\odot}$. 
An extrapolation of a typical EGRET spectrum 
up to $\sim$10~GeV is also shown (long-dashed line).}
\label{EGRET}
\end{figure}

Statistical studies of the EGRET sources in the galactic plane point to the association
between these sources and high density regions in the inner spiral arms including star
formation regions and molecular clouds (Romero et~al. \cite{Romero99}, Bhattacharya et~al.
\cite{Bhattacharya03}). It has been claimed that the low latitude EGRET sources can be
divided in two different subgroups, one formed by those sources that show variability, and
another formed by dubious or steady cases (Romero et~al. \cite{Romero04}). 
High-mass MQs have been proposed as possible variable EGRET source counterparts
(Kaufman Bernad\'o  et~al. \cite{kauf02}, Bosch-Ramon et~al. \cite{Bosch-Ramon04}). The
results of this paper show that MQs could be indirectly responsible for some of the 
non-variable EGRET sources. Taking into account that our model can reproduce a typical EGRET
spectrum (see Fig.~\ref{EGRET}) of a steady source, and the association between high-mass MQs
and high density regions, the scenario contemplated here could be the explanation of, at
least, a significant fraction of the non variable high-energy $\gamma$-ray detections. In
addition, these sources could produce detectable emission at TeV energies.

In Fig.~\ref{EGRET}, we show the predicted SED produced by the interaction of protons released
from a persistent jet of a MQ and hydrogen nuclei of a giant molecular cloud of
10$^5$~M$_{\odot}$ at $R$=10~pc and $t$=10$^6$~yr. At a typical Galactic distance, say
4~kpc, we obtain computed levels of emission in the EGRET range similar to the ones observed by
this instrument, with spectral slopes similar to the average value ($\sim$2.16) of this steady
subgroup of EGRET sources of the galactic plane (see Bosch-Ramon et~al. \cite{Bosch-Ramon04}).
These points make it reasonable to suggest an indirect association between the steady EGRET
population and MQs. We note that, due to the small distance between
the cloud and the MQ considered here in comparison to the cloud size, our results 
are a first order approach and an inhomogeneous model is required to improve 
the accuracy of our predictions, which will be
presented in a future work.

MQs can emit $\gamma$-rays by themselves (i.e. Bosch-Ramon et~al.
\cite{Bosch-Ramon04}). The $\gamma$-ray emitting MQ and cloud could have an angular  separation
too small to be distinguished by EGRET (although they could be distinguishable  at lower
energies). In such a case, the stronger emitter of $\gamma$-rays would dominate the spectrum and
variability properties and the next generation $\gamma$-ray instruments, with better angular
resolution than EGRET, would be required to separate them.

One interesting issue is the shape of the {\it young} proton spectra (see
Fig.~\ref{protesp}) and the subsequent electron energy distributions (see
Figs.~\ref{prel1}~and~\ref{prel2}). As is seen from the mentioned plots, there is a natural
low-energy cutoff in the proton spectrum produced by the energy-dependent diffusion, which
implies an almost monoenergetic distribution of electrons peaking at high energies. This
characteristic of the secondary particle energy distribution has a clear implication for the
related synchrotron spectrum: a slope like the one presented by a monoenergetic electron spectrum
at low frequencies with a spectral index of 1/3. Therefore, in the context of a MQ interacting
with a cloud, we expect to observe extended and steady emission with a spectral index of 1/3
at radio and even shorter wavelengths. For instance, detected radiation coming from the region
near the galactic center shows a similar spectrum  (see, i.e., Lesch et~al.
\cite{Lesch88}), possibly generated in a scenario like the one proposed here.

\section{Conclusions} \label{conclusions}

The study of the emission coming from high density regions of the ISM provides us with
information about accelerators of high energy particles located near these clouds. In this
work, we have explored the implications of the presence of a MQ close to a cloud. The obtained
results predict that extended radiation from radio frequencies to soft X-rays, generated by
synchrotron radiation, could be detected  from clouds, together with GeV and TeV $\gamma$-rays
produced mainly via $\pi^0$-decay. Further information can be obtained about cloud internal
characteristics (average magnetic fields, average densities). If the jets of
MQs are sources of relativistic hadrons, then nearby clouds can be indicators of such
hadronic accelerators with a specific broadband spectrum from radio to very-high energy
$\gamma$-rays.

\begin{acknowledgements}

We thank Diego Torres for useful comments. We also thank anonymous 
referee for constructive comments and suggestions.
V.B-R. and J.M.P. acknowledge partial support by DGI of the Ministerio de Ciencia y
Tecnolog{\'{\i}}a (Spain) under grant AYA-2001-3092, as well as additional support
from the European Regional Development Fund (ERDF/FEDER). During this work, V.B-R has been
supported by the DGI of the Ministerio de Ciencia y Tecnolog{\'{\i}}a (Spain) under the
fellowship FP-2001-2699.
\end{acknowledgements}

{}


\begin{thebibliography}{}

\bibitem[1996]{Aharonian&atoyan96} 
Aharonian, F.~A.~\& Atoyan, A.~M.\ 1996, \aap, 309, 917

\bibitem[2000]{Aharonian00} 
Aharonian, F.~A.~\& Atoyan, A.~M.\ 2000, \aap, 362, 937

\bibitem[2002]{Aharonian02} 
Aharonian, F. A., Belyanin, A. A., Derishev, E. V., Kocharovsky, V. V.,
 \& Kocharovsky, Vl. V. 2002, \prd, 66, 3005

\bibitem[1995]{Atoyan95} 
Atoyan, A.~M., Aharonian, F.~A., \& V{\" o}lk, H.~J.\ 1995, \prd, 52, 3265 

\bibitem[2003]{Bhattacharya03} 
Bhattacharya, D., Aky{\" u}z, A., Miyagi, T., Samimi, J., \& Zych, A.\ 2003, \aap, 404, 163 

\bibitem[1970]{Bg}
Blumenthal, G.~R.~\& Gould, R.~J.\ 1970, Reviews of Modern Physics, 42, 237 

\bibitem[2004]{Bosch-Ramon04}
Bosch-Ramon V., Romero, G.~E.~\& Paredes, J.~M.\ 2004, A\&A, in press,
[astro-ph/0405017]

\bibitem[2002]{Corbel02} 
Corbel, S., Fender, R.~P., Tzioumis, A.~K.; Tomsick, J.~A., et~al. 2002, Science, 298, 196

\bibitem[1964]{Ginzburg&syrovatskii64} 
Ginzburg, V.~L.~\& Syrovatskii, S.~I.\ 1964, The Origin of Cosmic Rays, New York: 
Macmillan, 1964

\bibitem[1999]{Gliozzi99} 
Gliozzi, M., Bodo, G., \& Ghisellini, G. 1999, MNRAS, 303, 37

\bibitem[2002]{Heinz02} 
Heinz, S.\ 2002, \aap, 388, L40 

\bibitem[2002]{Heinz&sunyaev02} Heinz, S.~\& Sunyaev, 
R.\ 2002, \aap, 390, 751 

\bibitem[2003]{Kaaret03} 
Kaaret, P., Corbel, S., Tomsick, J.~A., Fender, R., Miller, J.~M., Orosz, J.~A., Tzioumis, A.~K., 
\& Wijnands, R.\ 2003, \apj, 582, 945 

\bibitem[2002]{kauf02}
Kaufman Bernad\'o, M.~M., Romero, G.~E., \& Mirabel, I.~F. 2002, A\&A, 385,
L10--L13 

\bibitem[1988]{Lesch88} 
Lesch, H., Schlickeiser, R., \& Crusius, A.\ 1988, \aap, 200, L9 

\bibitem[2002]{Marshall02} 
Marshall, H.~L., Canizares, C.~R., \& Schulz, N.~S.\ 2002, \apj, 564, 941

\bibitem[2000]{Marti00} 
Mart{\'{\i}}, J., Paredes, J.~M., \& Peracaula, M.\ 2000, \apj, 545, 939

\bibitem[2003]{Meier03} 
Meier, D.~L.\ 2003, New Astronomy Review, 47, 667 

\bibitem[1994]{Mirabel&rodriguez94} 
Mirabel, I.~F.~\& Rodr{\'{\i}}guez, L.~F.\ 1994, \nat, 371, 4

\bibitem[1999]{Mirabel&rodriguez99} 
Mirabel, I.~F.~\& Rodr{\'{\i}}guez, L.~F.\ 1999, \araa, 37, 409

\bibitem[2003]{Nolan03}
Nolan, P.L, Tompkins, W.F., Grenier, I.A., Michelson, P.F. 2003, ApJ, 597, 615 

\bibitem[2000]{Paredes00} 
Paredes, J.~M., Mart{\'{\i}}, J., Rib{\'o}, M., \& Massi, M.\ 2000, Science, 288, 2340 

\bibitem[2002]{Paredes02}
Paredes, J.~M., Rib\'o, M., Ros, E., Mart{\'{\i}}, J., \& Massi,
M. 2002, A\&A, 393, L99

\bibitem[2002]{Ribo02}
Rib\'o, M. 2002, PhD Thesis, Universitat de Barcelona

\bibitem[1999]{Romero99}
Romero, G.~E., Benaglia, P., Torres, D.~F. 1999, A\&A, 348, 868

\bibitem[2003]{Romero03}
Romero, G.~E., Torres, D.~F., Kaufman  Bernad\'o, M.~M., \& Mirabel, I.~F. 2003, A\&A, 410, L1  

\bibitem[2004]{Romero04}
Romero, G.~E., Grenier, I.~A., Kaufman Bernad\'o, M.M., \& Mirabel, I.F., 
\& Torres, D.~F. 2004, ESA-SP, in press [astro-ph/0402285]

\bibitem[1984]{Spencer84} 
Spencer, R.~E.\ 1984, MNRAS, 209, 869

\bibitem[2003]{Tau03}
Tauris, T. M. \& van den Heuvel, E. 2003, Review to appear in the book 
'Compact Stellar X-Ray Sources', eds. W.H.G. Lewin and M. van der Klis, 
Cambridge University Press

\bibitem[2003]{Torres03}
Torres, D. F., Romero, G. E., Dame, T. M., Combi, J. A., \& Butt, Y. M. 2003, \physrep, 382,
303

\bibitem[2003]{Wang03} 
Wang, X.~Y., Dai, Z.~G., \& Lu, T.\ 2003, \apj, 592, 347 

\end{thebibliography}
\end{document}